\documentclass[sigconf]{acmart}

\AtBeginDocument{%
  }

\setcopyright{acmcopyright}
\copyrightyear{2022}
\acmYear{2022}
\acmDOI{XXXXXXX.XXXXXXX}

\acmConference[MM'22]{30th ACM International Conference on Multimedia}{October 10--14,
  2022}{Lisbon, Portugal}

\usepackage{booktabs}
\usepackage{url}
\usepackage{subfigure}
\usepackage{graphicx}
\begin{document}

\title{Bandwidth-Efficient Multi-video Prefetching for Short Video Streaming}

\author{Xutong Zuo}
\email{zuoxt18@mails.tsinghua.edu.cn}
\affiliation{
  \institution{Tsinghua University}
}

\author{Yishu Li}
\email{liyishu19@mails.tsinghua.edu.cn}
\affiliation{%
  \institution{Tsinghua University}
  }

\author{Mohan Xu}
\email{xu-mh19@mails.tsinghua.edu.cn}
\affiliation{%
  \institution{Tsinghua University}
  }

\author{Wei Tsang Ooi}
\email{ooiwt@comp.nus.edu.sg}
\affiliation{%
 \institution{National University of Singapore}
 }

\author{Jiangchuan Liu}
\email{jcliu@cs.sfu.ca}
\affiliation{%
  \institution{Simon Fraser University}
  }

\author{Junchen Jiang}
\email{junchenj@uchicago.edu}
\affiliation{%
  \institution{The University of Chicago}
  }

\author{Xinggong Zhang}
\email{zhangxg@pku.edu.cn}
\affiliation{%
  \institution{Peking University}
  }

\author{Kai Zheng}
\email{kai.zheng@huawei.com}
\affiliation{%
  \institution{Huawei Technologies}
  } 

\author{Yong Cui}
\email{cuiyong@tsinghua.edu.cn}
\affiliation{%
  \institution{Tsinghua University}
  }
\authornote{Corresponding authors}
\renewcommand{\shortauthors}{Trovato et al.}

\begin{abstract}
  Applications that allow sharing of user-created short videos exploded in popularity in recent years.  A typical short video application allows a user to swipe away the current video being watched and start watching the next video in a video queue.  Such user interface causes significant bandwidth waste if users frequently swipe a video away before finishing watching.  Solutions to reduce bandwidth waste without impairing the Quality of Experience (QoE) are needed. Solving the problem requires adaptively prefetching of short video chunks, which is challenging as the download strategy needs to match unknown user viewing behavior and network conditions. 
  In our work, we first formulate the problem of adaptive multi-video prefetching in short video streaming. 
  Then, to facilitate the integration and comparison of researchers' algorithms towards solving the problem, we design and implement a discrete-event simulator, which we release as open source.
  Finally, based on the organization of the Short Video Streaming Grand Challenge at ACM Multimedia 2022, we analyze and summarize the algorithms of the contestants, with the hope of promoting the research community towards addressing this problem.
\end{abstract}



\maketitle

\section{introduction} \label{sec:intro}

The popularity of user-generated, short-form, video sharing on social media platforms exploded in recent years. In 2020, Douyin and Kuaishou, two short video platforms, have more than 400 million and 300 million daily active users~\cite{zhang2022measurement} respectively.  Short video streaming, however, incurs, a high bandwidth cost.  
Kuaishou, for example, has an annual bandwidth cost of 5.7 billion RMB, accounting for 16.5\% of the total operational cost. 
In the first three quarters of 2021, the total bandwidth cost has exceeded that of the whole of 2020, showing an upward trend~\cite{kuai2021}. Therefore, saving bandwidth overhead without reducing user quality of experience (QoE) has become an important issue.

In short video applications, users watch videos in the mode of swiping and watching, i.e., a user may swipe away while watching the current video and start to watch the next video in the recommendation queue.  To ensure a smooth and high quality view experience, i.e., high QoE, the current video and videos in the recommendation queue need to be prefetched. However, if the user swipes away, the downloaded but unwatched video data no longer contribute to improving the user QoE, resulting in wasted bandwidth. As a result, videos need to be prefetched adaptively and strategically with the aim of reducing the bandwidth wastage when taking into account the user viewing behavior and network conditions.

\begin{figure*}[t]
    \centering
    \includegraphics[width= 0.9\textwidth]{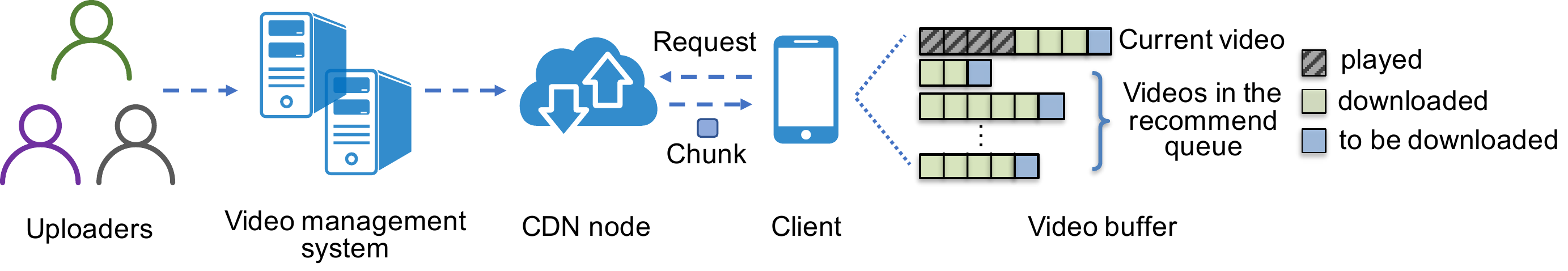}
    \caption{System overview for short video streaming.}
    \label{fig-overview}
\end{figure*}

Adaptive prefetching in the context of short video applications is challenging. Firstly, what to prefetch depends on the user viewing behavior, which is difficult to model as it is influenced by many factors, such as content and viewing history.  Secondly, there are multiple videos to prefetch, competing for the same share of bandwidth.  The player needs to consider which video to prefetch next.  Finally, the player also needs to decide the right bitrate to prefetch, using adaptive bitrate (ABR) algorithms~\cite{huang2014buffer,yin2015control,mao2017neural,yadav2017quetra}, adapting to changing network conditions.  In short video scenario, for the ABR algorithm, the video in the recommendation queue initially has no buffer. However, due to the existence of the previous video in the recommendation queue, there will be a ``virtual initial buffer'' for ABR, which will affect ABR decision.

Most earlier studies on reducing data wastage in video-on-demand focus on dynamic prefetching for a single video~\cite{chen2015smart,zhang2021post,chen2019energy}, which cannot be easily applicable to multi-video scenarios in short videos. Several recent works focused on prefetching for multiple short videos~\cite{plesca2016,he2020liveclip,zhang2020apl,lialfie,zhang2021short}. The industry currently either does not prefetch at all~\cite{zhang2020apl} (i.e., download the next video in queue only after the currently playing video is completely downloaded) or prefetches the video in a fixed order and a fixed size.  Both solutions can cause significant rebuffering and bandwidth wastage.  Later, dynamic prefetching is proposed~\cite{zhang2020apl} which converts the goal of minimizing bandwidth wastage to minimizing rebuffering when bandwidth is insufficient. However, the performance of bandwidth saving is not known when bandwidth is adequate. Zhang et al.~\cite{zhang2021short} leverages the results from the empirical study, that is, the long-term network conditions and viewing behaviors are highly predictable. They propose Wastage-Aware Streaming that learns viewing behaviors and network conditions to reduce data wastage while keeping QoE intact. It is, however, usually difficult to predict the user's viewing time. Leveraging the viewing behavior of different users shows the potential to prefetch video adaptively.

In this paper, we formulate the problem of adaptive multi-video prefetching in short video streaming (Section~\ref{sec:formulation}).  We analyze and present the optimization objective and its design space.
To address the challenges in comparing different prefetching algorithms and to provide a common platform for research by academics and industry, we design and implement an efficient short video streaming simulator (Section~\ref{sec:simu}), along with traces collected from a top-3 short video service in China.
Finally, to encourage the researchers from academia and industry to design adaptive prefetching solutions for short video streaming, we organize the Short Video Streaming Grand Challenge at ACM Multimedia 2022 (Section \ref{sec:challenge}) with the experience of holding two grand challenges before (Live Video Streaming Grand Challenge in 2019~\cite{yi2019acm} and Delay-Sensitive Video Streaming Grand Challenge in 2021~\cite{zhang2021acm}).
We analyze the submitted solutions and summarize their technical strengths and weaknesses in this paper.  We hope the platform, the dataset, and the insights gained from the grant challenge would spur development of solutions that will eventually be adopted and deployed in a real system.
\section{problem formulation} \label{sec:formulation}
Figure~\ref{fig-overview} shows the architecture of a typical short video streaming system.  Users create and upload short videos, which are then processed and encoded into different bitrate levels, and transmitted to the content delivery network (CDN) nodes. On the client side, users download and watch videos.  A user can swipe away the current video to start watching the next video in the recommendation queue.  In addition to the current video being played, the videos in the recommendation queue may also be downloaded (prefetched).  All videos are played in a certain order. The client works in a DASH~\cite{stockhammer2011dynamic} like fashion to decide the download chunk as well as its bitrate, and requests the desired chunks from the CDN node in a way that would maximize the QoE.

\subsection{Optimization Objective}

We add a new bandwidth penalty term to a widely-used QoE model~\cite{mao2017neural} as an optimization objective.  Furthermore, we separate the calculation of QoE terms for video playback and video download, since not all downloaded chunks are played.   Specifically, all downloaded chunks may cause rebuffering, and only chunks that are played need to be calculated for quality and smoothness. Consequently, the optimization objective is:

\begin{equation}
\begin{aligned}
    U_i =\;& QoE_i - Bandwidth_i \\
        =\ & \sum_{j} (\alpha \cdot R_j - \gamma \cdot S_j) -\sum_{k} \beta \cdot T_k - \sum_{k} \theta \cdot bw_k
\end{aligned}
\label{eq:qoe}
\end{equation}
where $j$ and $k$ count the chunks played and downloaded in video $i$ respectively. $R_j$ and $S_j$ measure the quality and smoothness for each played chunk $j$. $T_k$ and $bw_k$ are the rebuffering time and bandwidth usage caused by downloading chunk $k$.

The coefficients $\alpha$, $\beta$. $\gamma$, and $\theta$ are the weights for the bitrate, rebuffering time, smoothness, and bandwidth usage terms, respectively.  
In our simulator, we set $\alpha = \gamma = 1$, and $\theta = 0.5$ for simplicity. We set $\beta = 1.85$ to correspond to the highest bitrate level. 

\begin{table}[t]
\caption{Value range of decision variables.}\label{tab:DV}
\vspace{-0.2cm}
\begin{center}
\begin{tabular}{cc}
\toprule
\textbf{Decision Variables} & \textbf{Value Range}\\
\midrule
download video id & $[0,5$\\
bitrate & $\{750, 1200, 1850\}kbps$\\
sleep time (ms) & [0,$\infty$]\\
\bottomrule
\end{tabular}
\end{center}
\vspace{-0.4cm}
\end{table}

\subsection{Design Space}
To ensure user QoE in short video streaming, the current video and videos in the recommendation queue need to be prefetched. If the user swipes away, however, the downloaded but unwatched data are wasted. To achieve the optimization objective in Equation~\ref{eq:qoe}, the adaptive prefetching algorithm needs to decide which video to prefetch a chunk from at what bitrate, according to the network condition and the buffer state of the videos.  Our simulation allows prefetching from a maximum of 5 videos (including the current video being played back).

In addition, to reduce bandwidth wastage, prefetching may pause to reduce wastage when the network condition is good.  The sleep time should also be determined.  
The three decision variables used in our simulator (Section~\ref{sec:challenge} are listed in Table~\ref{tab:DV}. 
\section{short video streaming simulator} \label{sec:simu}

\begin{figure}[t]
    \centering
    \includegraphics[width= 0.45\textwidth]{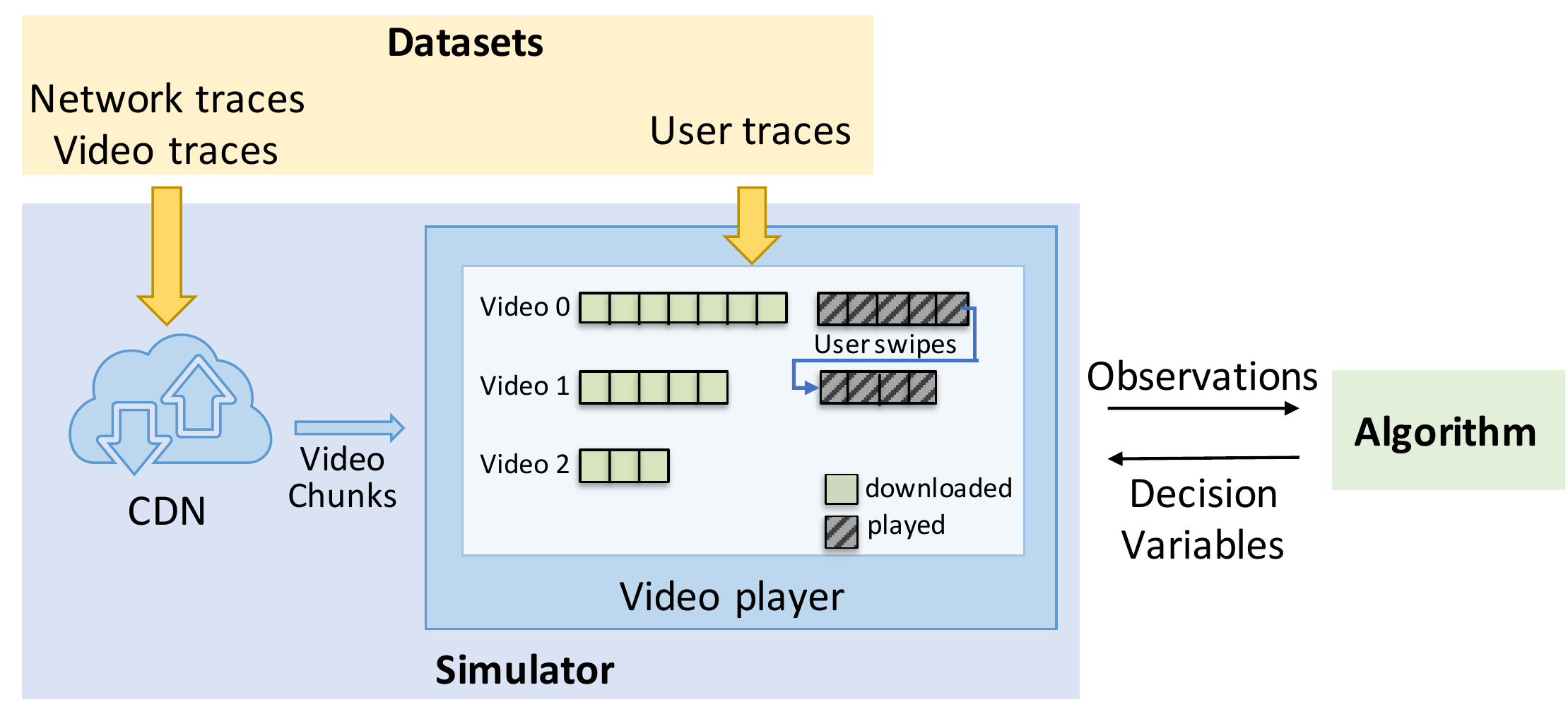}
    \vspace{-0.06in}
    \caption{Overview of short video streaming simulator.}
    \vspace{-0.14in}
    \label{fig-simu}
\end{figure}

To evaluate the adaptive prefetching algorithms, we provide a chunk-level simulator to model the workflow of short video streaming.
In the simulator, the video download process from CDN nodes and the playback of the current video are simulated.

\subsection{Architecture}
Figure~\ref{fig-simu} shows the overview of short video streaming simulator. 
The \textit{datasets} that are fed into the simulator contain network traces, video traces and user traces. 
The network traces record the bandwidth of the network to simulate a wide range of network conditions. The video traces describe the size of chunks in different bitrate levels. The network traces and video traces are required in the download process simulation.
The user traces provide users' viewing behavior, that is, users' watch time of each watched video is given. User traces are needed to simulate the video playback process. 
The \textit{algorithm}, which performs adaptive pre-rebuffering, interacts with the \textit{simulator} from which the algorithm get observations (listed in Table~\ref{tab:observations}). With these observations, the algorithm makes decisions and the output decision variables (Table~\ref{tab:DV}) guide the download process.

Due to the existence of multiple videos in the video player, the simulation is responsible for maintaining multiple buffers, differing from traditional video streaming that only maintains one video buffer. Specifically, for each downloaded chunk, the simulator calculates a download time according to the network bandwidth and the size of the chunk (parameter \textit{delay} in Table~\ref{tab:observations}). 
The buffer of current playing video is drained by the chunk download time and 
the playback duration (chunk length) is added to the video the downloaded chunk belongs to. 
The simulator keeps track of rebuffering events. Rebuffering happens when the buffer of the current playing video is empty. 
Rebuffering are more likely to occur when users swipe.
The rebuffering penalty is attributed to the chunk being downloaded when rebuffering occurs, representing by the parameter \textit{rebuf} in Table~\ref{tab:observations}.
To reduce bandwidth wastage, the simulator supports an essential operation: pause.  While pausing, the download process stops, but the playing process of the current video continues. 

\begin{table}[t]
\caption{Observations from Simulator.}\label{tab:observations}
\vspace{-0.1in}
\begin{center}
\begin{tabular}{cc}
\toprule
\textbf{Parameters} & \textbf{Description}\\
\midrule
delay & The time of pause or downloading a chunk\\
rebuf & Rebuffering time\\
video\_size & The size of the last downloaded chunk\\
end\_of\_video & Whether the last download video was ended\\
play\_video\_id & ID of the current video\\
Players & Data of videos in the players\\
first\_step & Whether it is the first operation\\
\bottomrule
\end{tabular}
\end{center}
\vspace{-0.1in}
\end{table}

\subsection{Implementation}
Based on the description above, we developed a short video streaming simulator for algorithm designers to evaluate their prefetching solutions.
In the implementation, the $network\_module$, $video\_module$ and $user\_module$ deal with the load and process network traces, video size traces and user behavior respectively. The $video\_player$ module executes the download and playback of multiple short videos. Algorithm designers are allowed to integrate their own modules, such as bandwidth estimation module, in the simulator to assist in decision-making. 
The simulator supports chunk-level download operations, which means that a chunk's download process cannot be stopped, and if a rebuffering event occurs during this period, the simulator will only download the expected chunk after the current download chunk is completed. 
For simplicity, the execution time of the simulator is accurate to millisecond and a smaller time cannot be executed with low practical significance.
We have open-sourced a version of the Python implementation of the simulator together with parts of the datasets~\cite{githubrepo2022}, with the hope of contributing to reproducible research.

\subsection{Dataset}
The dataset consists of three parts: network traces, video traces and user traces. Each application scenario is a combination of a network trace, a sequence of short videos, and a user's behavior watching these videos to simulate that a user watches a certain video sequence and applies a swiping operation under a controlled network condition.

\subsubsection{Network Traces}
The network trace records the network condition between a CDN node and the client. To facilitate the development of the algorithms, we provide a large scale of network traces. 
The public network traces~\cite{githubrepo2022} are synthetic and are constructed using the methods from Pensieve~\cite{mao2017neural}. Traces are divided into three categories, according to the average bandwidth: low, medium and high, representing the weak, medium and excellent network conditions, using the threshold of 1.5 and 3~Mbps respectively. Participants can design, evaluate and refine their algorithms with these traces, or use the same method to construct more network traces to use. The network traces used for evaluation are randomly selected from the collected real network traces in WiFi and 4G scenarios provided by \textit{PowerInfo}\footnote
{http://www.powerinfo.net/.}. These traces are also divided into three categories with the division boundaries 1.9 and 3~Mbps with the consideration of the video highest bitrate (1.85~Mbps). Each network trace is a text file containing multiple lines. Each line contains two floating-point numbers: the timestamp in seconds and the measured throughput in Mbps. 

\begin{table}[t]
\caption{Description of videos.}\label{tab:videos}
\vspace{-0.1in}
\begin{center}
\begin{tabular}{ccc}
\toprule
\textbf{Video Name} & \textbf{Length(s)} & \textbf{Video Type}\\
\midrule
tj & 17 & Education\\
EDG & 26 & Entertainment\\
gy & 37 & Campus Life\\
dx & 40 & Campus Life\\
ss & 47 & Campus Life\\
jt & 6 & Entertainment\\
yd & 125 & Game\\
\bottomrule
\end{tabular}
\end{center}
\vspace{-0.3cm}
\end{table}

\subsubsection{Video Traces}
The video traces contain chunk-level traces of video sequences. There are seven videos open-sourced\cite{githubrepo2022} as an example for the algorithm designers, covering the video type of \textit{Education}, \textit{Entertainment}, \textit{Campus Life} and \textit{Game} with various video length (shown in Table~\ref{tab:videos}).
The video traces are collected from a top-3 short video service in China. The original videos are 720p, and we re-encode the high quality videos into three resolution \{360,480,720\}p of bitrate at \{750,1200,1850\}~kbps with a H.264/MPEG-4 codec. After that, we divided the video into 1-second chunks. The traces are stored in text format, with each line corresponding to the size of a chunk in a certain video representation. 

\subsubsection{User Traces}
User traces describe a user retention model of each video, reflecting the probability of users leaving at different times.
The user retention of a video is the statistics of all users who watched the video. 
It describes the percentage of users left after each second. For example, a video of 3 seconds: [(0, 1), (1,0.9298), (2,0.8324), (3,0.7298), (4,0)] with the form (timestamp, percentage of left users).
It means that 92.98\% of users are still watching at the end of the first second, 83.24\% of users are still watching at time 2.0s. Specifically, 72.98\% of users still exist at 3.0s, which means that they almost watched the whole video. The last line (4,0) is an end mark.

\section{Grand Challenge Summary} \label{sec:challenge}

\subsection{Summary of Submissions}

In this grand challenge, 139 teams from both academia and industry have registered. 65 teams have submitted their algorithms and finally 27 teams completed all the competition process and obtained the corresponding ranking. 
8 of the top 10 teams use heuristic algorithms and 2 adopt learning-based algorithms. They leveraged the bandwidth prediction, the buffer of each video and the user retention rate to determine which video to download and its bitrate. Buffer thresholds are set, the upper threshold used to keep the buffer of each video from being too large, and the lower threshold used to avoid too many rebuffering events.

All submitted heuristic algorithms utilized a bandwidth estimation scheme to assist in decision-making. Some of them used an existing scheme, such as harmonic average and moving average. Some teams adjust the parameters of their algorithms based on the bandwidth prediction to change how conservative the algorithm is, or switch to different ABR algorithms under different network conditions.Some submitted learning-based solutions used actor-critic framework~\cite{konda1999actor} in reinforcement learning and learned from past streaming experiences. Among the top 10 teams, the two learning-based algorithms are placed at 8th and 10th. The performance of learning-based schemes may be affected by the lack of generalization ability when faced with various combinations of different network conditions, video sequences, and user behaviors.  The number of public traces provided may not be adequate for training, leading to overfitting.

\subsection{Evaluation and Ranking}

For a submitted algorithm, we evaluated it in the simulator with the combinations of network traces, video traces, and user traces. Specifically, we run the algorithm under each condition (a video sequence, a network trace, and a user viewing behavior) and get the score. 
Then we normalize the scores of all contestants under this condition to eliminate the difference in scale of different evaluation conditions. Specifically, we calculate $Normal\_S$ = $(S-MAX)/(MAX-MIN)$ where MAX is the maximum scores of all contestants under this condition and MIN is the score of the baseline algorithm we implemented. The final score of an algorithm is the sum of $Normalized\_S$ under all conditions.  The teams are ranked according to the sum.  The rank of a team is the highest ranking of all submitted algorithms for this team in the same evaluation phase. 

\begin{figure} 
\centering
\subfigure[Sample 25 times]{\includegraphics[width=0.32\linewidth]{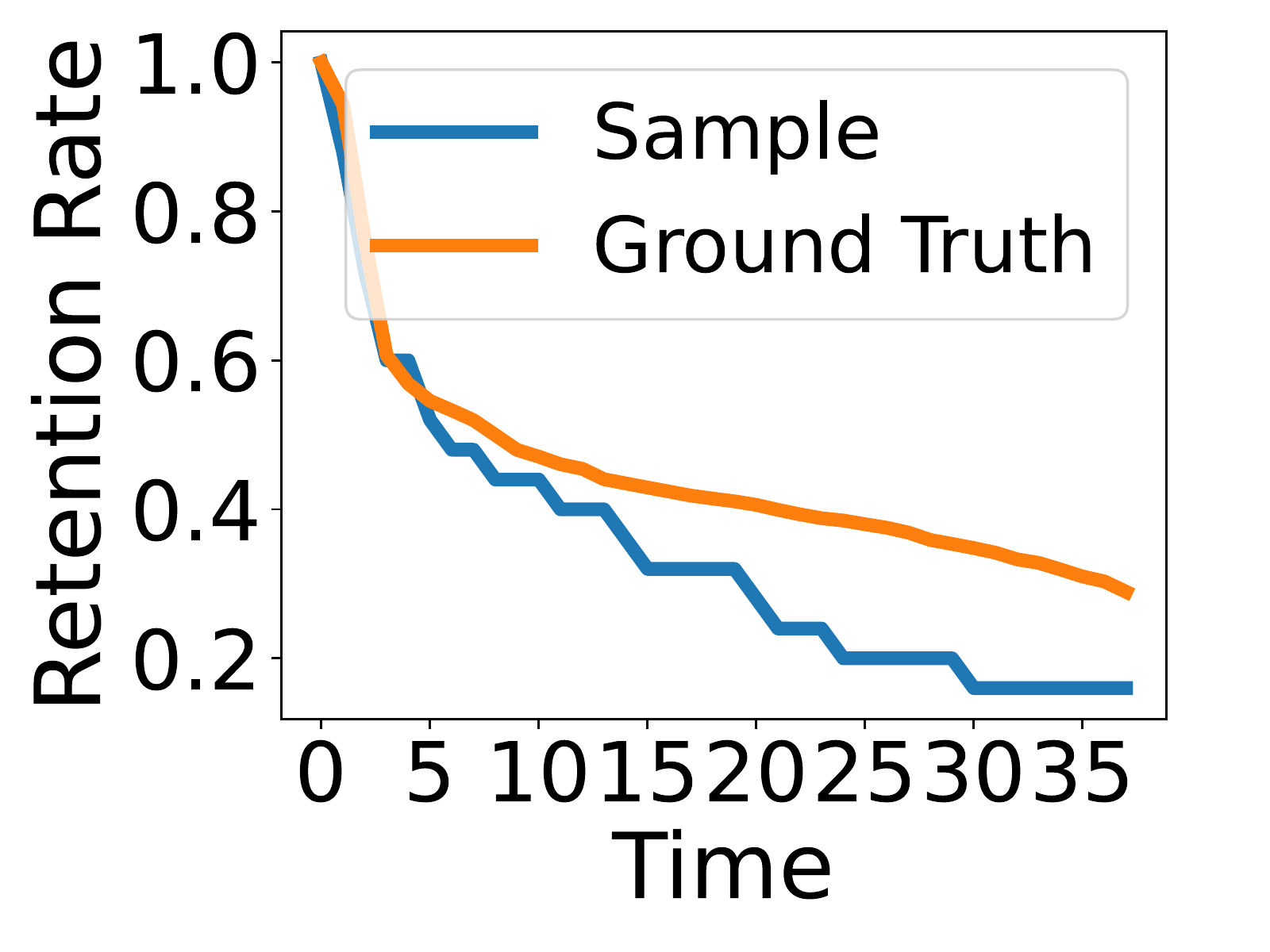}}
\subfigure[Sample 50 times]{\includegraphics[width=0.32\linewidth]{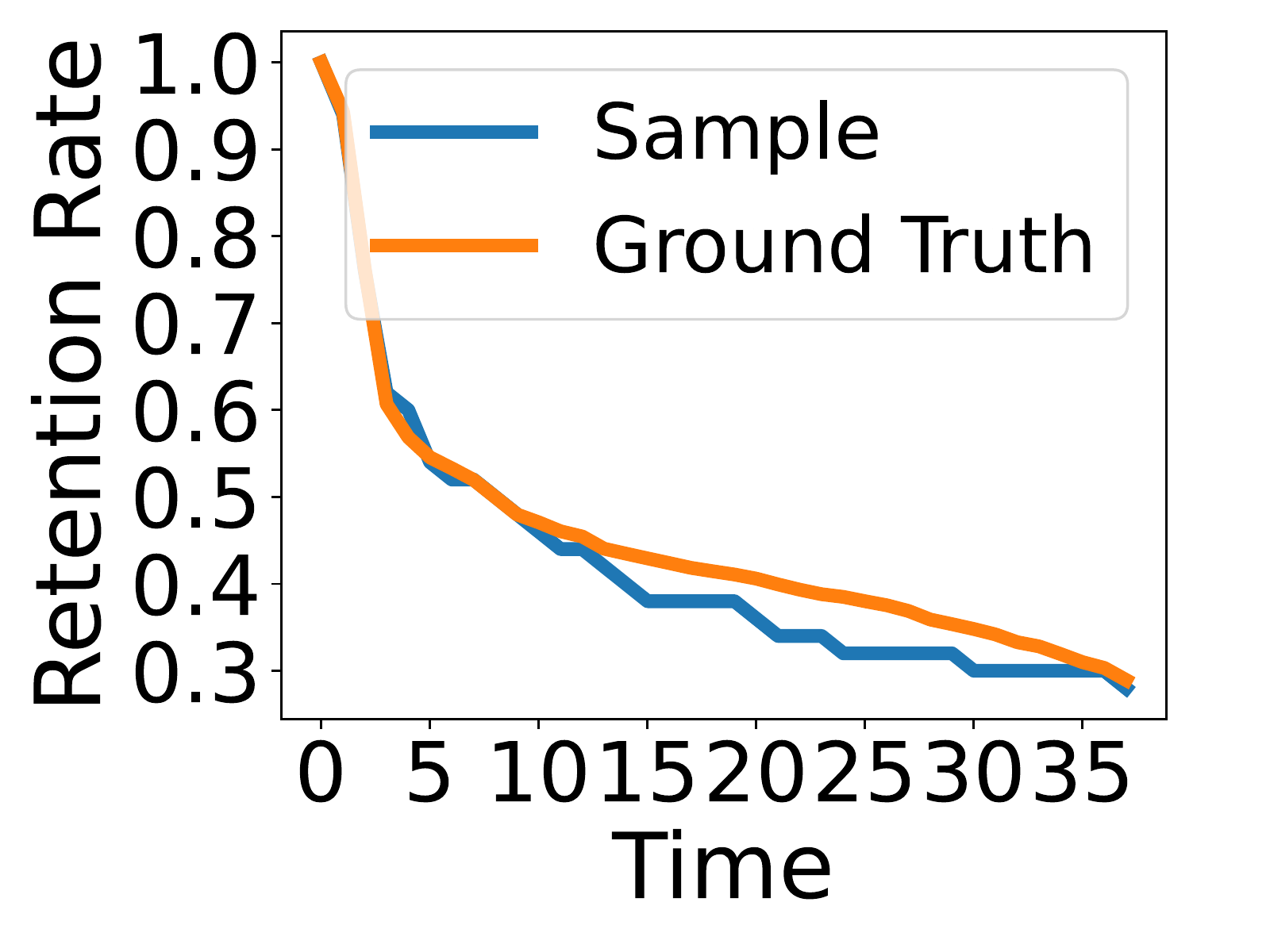}}
\subfigure[Sample 100 times]{\includegraphics[width=0.32\linewidth]{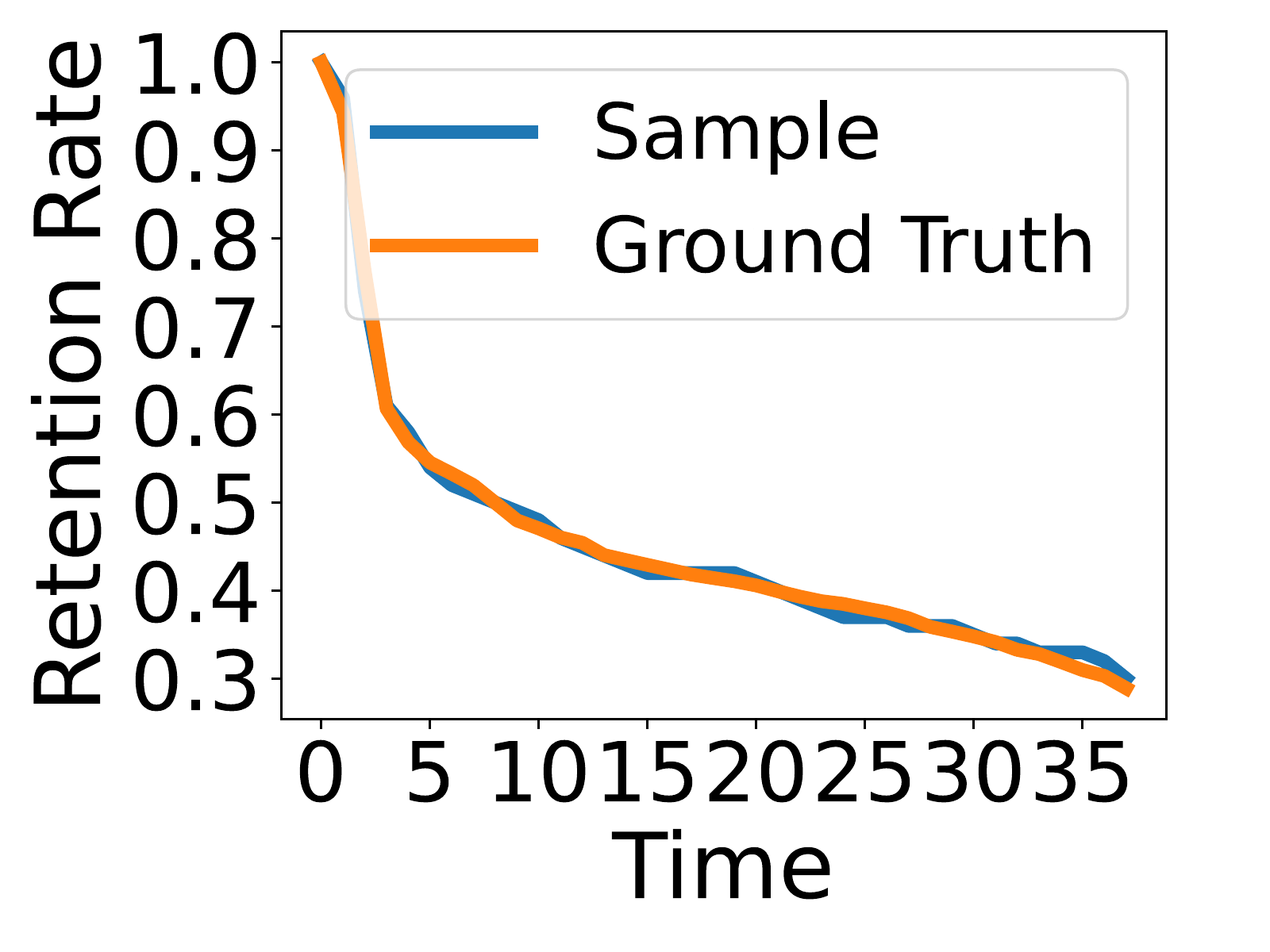}}
\vspace{-0.1in}
\caption{User retention curves with different sample times.}
\label{fig:sample}
\vspace{-0.2in}
\end{figure}

During evaluation, we discovered that the number of samples on the user retention curve is an important parameter. 
Randomness increases as the number decreases; the evaluation time increases with the number increases.
We compare the user retention rate curve obtained by sampling $N$ times for the viewing duration of $N$ users watching the same video sequence with the real retention rate curve, as shown in the Figure~\ref{fig:sample}. This is an example video sequence.  Taking $N$ as 25, 50, 100, we find that, as the number of samples increases, the curve fits the real curve better as expected.
However, with multiple network conditions to be evaluated, too many user samples increase the number of evaluation conditions and thus increase the evaluation time. After weighing the fitting results and evaluation time, we choose $N=50$ where the evaluation time of an algorithm is several minutes, which is acceptable.

\section{conclusion}

We formulated the problem of adaptive multi-video prefetching in short video streaming and provided a simulation platform that allows researchers to evaluate the proposed algorithms for short video streaming. We are grateful that researchers participated in ACM Multimedia 2022 Short Video Streaming Grand Challenge and contributed excellent algorithms to our community. We hope that the designed algorithms can finally be applied to the industry.

\balance

\bibliographystyle{ACM-Reference-Format}
\bibliography{refer}

\appendix

\end{document}